# Thermal strain-induced enhancement of electromagnetic properties in SiC-MgB$_2$ composites

R. Zeng[1] S.X. Dou[1,a)], L. Lu[1], W. X. Li[1], J.H. Kim[1], P. Munroe[2], R.K. Zheng[3] and S.P. Ringer[3]

[1]Institute for Superconducting and Electronic Materials, University of Wollongong, Northfields Avenue, Wollongong, NSW 2522, Australia

[2]Electron Microscope Unit, University of New South Wales, Sydney, NSW 2052 Australia

[3]Electron Microscope Unit, University of Sydney, Sydney, NSW 2000 Australia

a) Corresponding author's e- mail: shi@uow.edu.au

Strain engineering has been used to modify materials properties in ferroelectric[1-3], superconducting[4-7], and ferromagnetic thin films[8]. The advantage of strain engineering is that it can achieve unexpected enhancement in certain properties, such as an increase in ferroelectric critical temperature, $T_c$, by 300 to 500°C, with a minimum detrimental effect on the intrinsic properties of the material[1,2]. The strain engineering has been largely applied to the materials in thin film form, where the strain is generated as a result of lattice mismatch between the substrate and component film or between layers in multilayer structures. Here, we report the observation of residual thermal stress/strain in dense SiC-MgB$_2$ superconductor composites prepared by a diffusion method. We demonstrate that the thermal strain caused by the different thermal expansion coefficients (α) between the MgB$_2$ and SiC phases is responsible for the significant improvement in the critical current density, $J_c$, the irreversibility field, $H_{irr}$, and the upper critical field, $H_{c2}$, in the SiC-MgB$_2$ composite where the carbon substitution level is low. In contrast to the common practice of improving the $J_c$ and $H_{c2}$ of MgB$_2$ through chemical substitution, by taking advantage of residual thermal strains we are able to design a composite, which shows only a small drop in $T_c$ and little increase in resistivity, but a significant improvement over the $J_c$ and $H_{c2}$ of MgB$_2$. The present findings open up a new direction for manipulation of materials properties through strain engineering for materials in various forms.

The doping of nano-SiC particles to MgB$_2$ has been proven to be particularly effective in significantly enhancing the $J_c$, $H_{irr}$ and $H_{c2}$.[9-12] A record $H_{irr}$ of 29 T and $H_{c2}$ (4.2 K) of 43 T has been achieved with nano-SiC doped MgB$_2$[11,13]. The in-field $J_c$s of the nanoparticle doped MgB$_2$ increased by more than an order of magnitude. These results have been confirmed by many groups around the world[14-16]. The most recent report shows that $J_c$s of 4.1×10$^4$ A/cm$^2$ at 4.2 K and 10 T has been achieved with nano-SiC doped MgB$_2$ wires[17]. Extensive work has been carried out to understand the mechanism that explains the special doping effects of C and SiC[10-12,18-19]. It has been demonstrated that C substitution into B sites can introduce electron scattering[19], which reduces the mean free path and the coherence length ($\xi$), increases the residual resistivity ($\rho(40 K)$), and enhances $H_{c2}$. In a previous work we have demonstrated that the particle size and crystallinity of the SiC is very critical to the enhancement of $J_c$ and $H_{c2}$ by SiC doping, that is, only doping via nano-sized SiC facilitated a significant increase in $J_c$ and $H_{c2}$, while micron-scale crystalline SiC doping through an *in-situ* reaction process had little effect, since it did not result in C substitution[20]. Recently, a dual reaction model has been proposed to provide a more comprehensive understanding of nano-SiC doping in MgB$_2$[2]. According to the dual reaction model, the nano-SiC doping allows C substitution and MgB$_2$ formation to take place simultaneously at low temperatures. The dual reaction model not only explains the unique features of nano-SiC doping in MgB$_2$, but also provides a means of comprehensive assessment for many other dopants, as well as predicting other dopants suitable for enhancing the performance properties of MgB$_2$. The disadvantages of nano-SiC doping are the reduction in $T_c$ and the increase in resistivity, thus resulting in deterioration of low-field $J_c$.

In this letter, we propose an alternative route to enhance the electromagnetic properties by strain engineering. We designed and fabricated a SiC-MgB$_2$ composite by a diffusion reaction of Mg with a premixed B-SiC bulk. We found that all of the samples showed much higher purity and higher density than those of *in-situ* mixed powder reaction bulks. The SiC-MgB$_2$ composite sample showed a significant enhancement of $J_c$ and $H_{c2}$. Such prepared samples have little carbon substitution and show very small changes in critical temperature ($T_c$), and resistivity $\rho(40\ K)$. Furthermore, thermal strain analysis indicated that the different thermal expansion coefficients ($\alpha$) of $MgB_2$ and SiC introduce strains into the $MgB_2$ matrix surrounding the SiC particles during the cooling process. These strains caused a huge stress field and many defects in the $MgB_2$ matrix, which were confirmed by XRD, TEM, Raman measurements and $J_c$ temperature dependence analysis, and these are believed to be responsible for the improvement in $J_c$ and $H_{c2}$.

Starting powders of crystalline B 99.999%, with and without 10 wt% SiC particles, were mixed and pressed into pellets. The pellets were then put into an iron tube filled with Mg powder (99.8%). The atomic ratio between Mg and B was 1.15:2.0. Since the formation of $MgB_2$ was accomplished via a diffusion process, long reaction times were thus needed to obtain a fully reacted $MgB_2$ bulk. The samples were sintered at 923 K–1223 K for 10 hrs in a quartz tube, where a flow of high purity argon gas was maintained, and then furnace-cooled to room temperature. All samples were characterized by X-ray diffraction (XRD) and analyzed using Rietveld refinement XRD to determine the *a* and *c* lattice parameters and the MgO content. Microstructural observations were performed by using a scanning electron microscope (SEM), field-emission SEM, and a transmission electron microscope (TEM). The $T_c$ was defined as the onset temperature at which diamagnetic properties were observed. The Raman scattering was measured by a confocal laser Raman spectrometer (Renishaw inVia plus) with a 100× microscope. The 632.5nm line of an $Ar^+$ laser was used for excitation, with the laser power maintained at about 20mW, measured on the laser spot on the samples, in order to avoid laser heating effects on the studied materials. The magnetization of samples was measured at 5 and 20 K using a Quantum Design Physical Properties Measurement System (PPMS) with a magnetic field sweep rate of 50 Oe /s and amplitude up to 9 T. The magnetic $J_c$ was calculated from the height of the magnetization loop, M, using the critical state model: $J_c = 12Mb/d(3b-d)$, with b and d the dimensions of the samples perpendicular to the direction of applied magnetic field and d < b. The magnetoresistivity $\rho(H,T)$ was measured with $H$ applied perpendicular to the current direction, using the four probe method in the temperature range from 4.2 K to 300 K and the field range from 0 T to 9 T. The irreversibility field, $H_{irr}$, and $H_{c2}$ could be deduced using the criteria of 0.1 and 0.9 of $\rho(H,T)$, respectively.

Fig. 1 shows the Rietveld refinement XRD patterns of the pure $MgB_2$, the 10wt% $SiC-MgB_2$ composite samples, and XRD patterns of SiC particles added in $SiC-MgB_2$ composite. By using the Rietveld refinement analysis, the $a$- and $c$-axis lattice parameters and SiC content were determined as shown in Table I, which also gives a comparison of the density, the defect induced $a$- and, $c$-axis lattice strain, the grain size, the percentage of Mg vacancies, $T_c$, and residual resistance at 40K ($\rho(40K)$) for the two samples. We note that the $a$-axis parameter is virtually the same for both the doped and un-doped samples, while $c$-axis parameter is slightly enlarged in the $SiC-MgB_2$ composite. In contrast, the $a$-axis parameter for *in situ* processed SiC doped $MgB_2$ is reduced while $c$-axis parameter should remain unchanged, as reported by a number of groups[9-13]. It is also interesting to note that the SiC particles remained un-reacted and formed a composite with the $MgB_2$ in the SiC-doped sample. The Rietveld refinement analysis results showed that there was about 9.3wt% SiC, which is about the same as in the starting precursor, this may due to both the crystalline not amorphous of SiC and the diffusion process. This is consistent with the fact that the XRD pattern showed no presence of $Mg_2Si$ (Fig. 1), This is in clear contrast to the SiC-doped $MgB_2$ prepared by the *in-situ* technique[10-12], in which very small amount of SiC remained while $Mg_2Si$ is always present due to the reaction of Mg with SiC.

All these samples appeared to exhibit a high density (about 80% of the theoretical density, compared with about 50% of the theoretical density for *in-situ* mixed bulk). There is a small drop (0.6 K) in $T_c$ and little increase in $\rho(40\ K)$ (from 12 $\mu\Omega$ cm to 16 $\mu\Omega$ cm) for the SiC-doped sample. In contrast, the *in-situ* processed SiC doped $MgB_2$ normally had a decrease in $T_c$ by 1.5 to 2 K[9] and an increase in $\rho(40\ K)$ from 90 $\mu\Omega$ cm to 300 $\mu\Omega$ cm[12]. These $SiC-MgB_2$ composite samples had a high $T_c$ value (37.8 K), and low $\rho(40\ K)$ (Table I). From all the data presented, we would expect that there would be no significant effect of SiC doping on $J_c$ and $H_{c2}$, since there is very low level of C substitution, if any, for B and little increase in resistivity. However, it is interesting to note that the SiC-MgB$_2$ composite sample showed not only an improved in-field $J_c$ but also no degradation in self-field $J_c$, as shown in Fig. 2 and Table I, while it has been well established that nano-SiC doping using *in-situ* technique reduces the self-field $J_c$[1,5]. A $J_c$ of 5.8 × 10$^5$ in self-field and 1.9 × 10$^4$ in 4 T at 20K for SiC-MgB$_2$ composite can be compared to the $J_c$ of 4.1 × 10$^5$ in self-field and 1.4 × 10$^3$ in 4 T at 20K for pure MgB$_2$. Fig. 3 shows $H_{irr}$ and $H_{c2}$ as a function of the temperature for pure and SiC-MgB$_2$ composite samples. As can be seen, $H_{irr}$ and $H_{c2}$ of the SiC-MgB$_2$ sample are significantly improved in comparison with the pure sample.

The magnitude of enhancement in terms of $J_c$, $H_{irr}$, and $H_{c2}$ for the SiC-MgB$_2$ composite prepared by the diffusion method in this work is comparable to the values of these properties for nano-SiC doped MgB$_2$ prepared by the reaction *in-situ* method. More important is the improvement of the low-field $J_c$ for the SiC doped sample, which always showed deterioration, as in most doped MgB$_2$, due to the reduction in the effective cross section of the superconductor phase. These results raise a serious question on the nature of the mechanism behind the significant property enhancement in SiC-MgB$_2$ composite, as these effects can not be explained by the dual reaction model[10] or C substitution[12]. It is clear that SiC inclusions in the SiC-MgB$_2$ composite, with little C substitution, can evidently induce strong pinning and scattering in MgB$_2$, so as to improve $J_c$ and $H_{c2}$. Because the residual SiC is the dominating impurity, this enabled us to single out the special effects of residual SiC among the many impurities which are normally present in samples made using the reaction *in situ* process. As the difference in thermal expansion coefficients ($\alpha$) between SiC and MgB$_2$ is quite large, which can create thermal strains in the MgB$_2$ matrix during cooling we will examine whether the residual thermal strain could be an effective pinning mechanism.

It is reasonable to assume that at the MgB$_2$ formation temperature of 1123 K, both the MgB$_2$ and the SiC are in a stress-free state, due to the relatively high sintering temperature over a long period of time. Furthermore, thermal expansion of crystalline solids can be attributed to the anharmonicity of the interatomic potential and therefore considered to be a property intrinsic to each material. Since the thermal expansion coefficient, $\alpha$, of SiC is smaller than that of $MgB_2$, the difference in $\alpha$ between $MgB_2$ and SiC dictates the stress status of the $MgB_2$ and causes thermal strain in the $MgB_2$ matrix around the SiC particles during the cooling process. Fig. 4 shows plots of the thermal expansion coefficient, $\alpha$, for $MgB_2$ and SiC along the *a*- and *c*-axes (Fig. 4(a)), using data taken from the literature [21-24], from which the normalized lattice change and lattice strain in the $MgB_2$ matrix during cooling from 1123K to 0K can be derived, as shown in Fig. 4(b). It should be noted that $\alpha$ for $MgB_2$ is much larger than for SiC. Furthermore, the former is characterized by high anisotropy, while the latter is nearly isotropic. If the two phases are strongly bonded, the $MgB_2$ phase will be subjected to a tensile strain on cooling, in particular, and there will be a larger strain along the *c*-axis, since there is a larger difference in $\alpha$ along the *c*-axis between $MgB_2$ and SiC. The normalized lattice strain is as large as – 0.55% in the doped $MgB_2$ along the *c*-axis at room temperature. The negative value corresponds to tensile strain in the $MgB_2$. This value is about equal to the critical strain rate measured for multifilament $MgB_2$ wires[25], which showed that the $J_c$ increased with increasing tensile strain rate, although this strain was measured on the macro-scale. The large *c*-axis strain in the doped $MgB_2$ resulted in an enlargement in the *c*-axis by 0.15 % in comparison with pure $MgB_2$, as shown in Table I. The increased strain in the SiC-$MgB_2$ is also evident from the larger full width at half maximum (FWHM) values (0.48° for the 002 peak) than for pure $MgB_2$ (0.326° for the 002 peak). Since the $MgB_2$ grain size is the same in both SiC-$MgB_2$ and pure $MgB_2$ (Table 1 as observed by TEM examination) the increase in the FWHM can be solely attributed to the increase of strain. Using Williamson-hall analysis, we calculated lattice strain to be 0.208 and 0.306 in *a*-axis, and 0.292 and 1.13 in *c*-axis for pure and SiC-$MgB_2$ respectively (Table 1). The lattice strain in *c*-axis in the SiC-

MgB$_2$ increased from pure MgB$_2$ by a factor of 4, attributable to the high anisotropy in thermal expansion coefficient of MgB$_2$.

The thermal strain could create a huge stress field, structural defects, and lattice distortion in the MgB$_2$ matrix, which would remain in the SiC-MgB$_2$ sample during cooling. TEM examination revealed the following features: (1) SiC and MgB$_2$ remain as separate phases and form a strongly bonded composite, as shown in Fig. 5(a). The SiC particles, whilst micron sized in scale, are comprised of smaller grains a few nanometers in size. It is these small grains which assist in providing excellent wetting and contact with the MgB$_2$ phase. (2) The grain size for both pure MgB$_2$ and MgB$_2$ in the SiC-MgB$_2$ composite is the same (~100 nm). (3) There is a high density of defects (dislocations and lattice distortion) in the MgB$_2$ phase along the interfaces between the SiC and MgB$_2$, as indicated by the arrows in Fig. 5(a), which may be attributable to the tensile strain in the MgB$_2$ phase that is imposed by the small thermal expansion of SiC during cooling. Fig. 5(b) shows the interface between the two coexisting phases: MgB$_2$ and SiC. The electron diffraction patterns taken from both sides indicate that the two phases are faceted with [101] of SiC and [001] of MgB$_2$. The thermal expansion coefficient for MgB$_2$ is highly anisotropic with a large variation in [001] direction while that for SiC is nearly isotropic. Thus, this kind of interfaces will impose tensile stress along *c* axis in MgB$_2$. As a result, a lot of wave fringes were observed in MgB$_2$ near the interface of SiC as shown in Fig. 5(c) and Fig. (d). In comparison, these wave structures are hardly seen in the pure MgB$_2$ sample. Careful examination indicated that these fringes were induced by two layers lattice mismatch, so dislocations and lattice distortions were commonly observed in fringes area. The lattice distortion observed in TEM confirmed the Raman measurements results described below.

Raman spectroscopy is known to be an excellent probe for the detection and estimation of these stresses and strains, or lattice distortions, since the electron-phonon coupling intensity and crystal distortion will influence the Raman shift and the line-width of the Raman scattering, which can give some insight into the anharmonicity/harmonicity competition of the $E_{2g}$ mode. The peaks centered at the phonon density of states (PDOS) are understood to arise due to disorder and distortion[26-28]. A compressive stress is characterized by a shift of the Raman peak to higher frequencies, while a tensile stress is characterized by a shift of the Raman peak to lower frequencies[29]. The relative intensity and FWHM of the PDOS peaks reflect the extent of the distortion[30]. Fig. 6 shows the normalized ambient Raman spectra of the samples sintered at 1123 K for 10 hrs, with the line and dot spectra corresponding to pure $MgB_2$ (black) and SiC doped $MgB_2$ (red), before and after cooling to low temperature, respectively. There are four peaks in the measurement range from 200cm$^{-1}$ to 1000cm$^{-1}$, centered at about 300cm$^{-1}$, 400cm$^{-1}$, 600cm$^{-1}$ and 770cm$^{-1}$. In this study, we focus on the two peaks centered at 600cm$^{-1}$ and 770cm$^{-1}$, since the broadened peak at 600cm$^{-1}$ is associated with the $E_{2g}$ mode, while the peaks at 400 cm$^{-1}$ and 770 cm$^{-1}$ are associated with the PDOS. The peak at 300cm$^{-1}$ has not been reported in the literature so far. The $E_{2g}$ and PDOS peaks are shifted to lower frequency for the SiC-doped sample ($E_{2g}$: 576cm$^{-1}$, PDOS: 762cm$^{-1}$) compared with the pure sample ($E_{2g}$: 600cm$^{-1}$, PDOS: 770cm$^{-1}$), which indicates that there is residual tensile strain in the original SiC-$MgB_2$ composite sample. By comparing the Raman spectra of samples before and after cooling to low temperature (down to 10 K), we find that the most obvious changes in the Raman peaks are that the shifting tendency of the two main Raman peaks, the FWHM of both peaks, and the intensity of the main PDOS peak are increased by the cooling process for the SiC-doped sample, but not changed for the pure sample. The enhanced PDOS peak in the SiC-doped sample is an indication of the increased residual strain, which leads to enhanced flux pinning and thus improves the $J_c$ and $H_{c2}$. When the α of SiC particle is smaller than the α of $MgB_2$, this will cause tensile strain in the $MgB_2$ matrix around the dopant particles. However, it must be pointed out that the difference in α of the two phases alone can not guarantee the occurrence of large strain. The large strain status can only be generated if the two phases are strongly bonded and the density of the composite is high enough. As reported previously, micron-size SiC particle doping into $MgB_2$ using an *in-situ* reaction process did not give rise to any improvement in $J_c$ or $H_{c2}$[16]. The reason for this is that the *in-situ* reaction process results in more than 50% porosity in the SiC-doped $MgB_2$. The high porosity can easily accommodate the volume expansion and relax the strain between the SiC and the $MgB_2$ matrix. Furthermore, in reaction *in situ* case SiC is largely reacted through reaction: $SiC + Mg = Mg_2Si + C$. According to the dual reaction model, simultaneous reaction between the dopant and matrix components is essential for effective doping. Although overall C substitution level is very low as indicated by the XRD data, the reaction between the SiC particles and the parent precursor at the interface enables wetting and strong bonding between residual SiC and $MgB_2$. It is believed that there may be some localized C substitution in $MgB_2$ in the vicinity of interfaces. The conditions for the creation of thermal strain-induced enhancement of $J_c$, $H_{irr}$, and $H_{c2}$ include (i) a large difference in $\alpha$ for the two components in the composite, (ii) good compatibility and strong bonding between the two phases at interfaces, (iii) lattice mismatch and (iv) a high density.

The thermal strain seems to be effective in increasing correlated linear defects, i.e., high density dislocation lines, stacking defects, etc., and produces uncorrelated disorder too, i.e., a random distribution of defects, e.g., globular non-superconducting regions on the scale of the superconducting coherence length ($\xi$), which thus enhances flux pinning. Additional pinning in $MgB_2$ may originate from stress/strain field centers. In element substituted (such as C substituted) and Mg deficient $MgB_2$ additional stresses are created by the locally disturbed lattice, i.e., lattice misfit. Since the spatial distribution of stress fields is attenuated as $1/r^6$ ($r$ is the radius of the spherical imperfection or stress center), their range of influence is rather localized, being comparable with $\xi$. We assume that stress/strain fields induced by lattice misfit are randomly distributed in the $SiC + MgB_2$ composite, contributing to flux pinning in a collective behavior, leading to higher $J_c$, and an upwards shift of the $H_{irr}$. The contribution may become more remarkable for vortex matter in the regime of single vortex and vortex-vortex interaction. Note that the enhanced pinning from Mg vacancies can be overlooked, as both the pure $MgB_2$ and the SiC + $MgB_2$ composite presented here are the same in this respect (3.2% and 3.6%, as shown in Table I), having similar $T_c$s as high as 38.4 K. To further evidence and understand the stress-strain induced pinning mechanism, we return to analyzing the experimental temperature dependence of $J_c$. According to the model of thermally activated flux motion and the model of collective flux pinning[37-42], flux pinning mechanisms are mainly classified into two or three types: (i) $\delta T_c$, due to the spatial fluctuation of the Ginsburg-Landau coefficient associated with the transition temperature $T_c$, (ii) $\delta l$, which is induced by the spatial fluctuation of the effective mass related to the charge carrier mean free path $l$, or (iii) $\delta\varepsilon$, which is due to the stress/strain field. The temperature dependence of the critical current density $J_c(T)$ can be expressed as:

$$J_c(T) \approx J_c(0) f(t)$$

$$\text{with } f(t) = (1-t^2)^X (1+t^2)^Y$$

where $J_c(0)$ is $J_c$ at the temperature of 0 K, and $t$ is the normalized temperature $t = T / T_c$. The association with the different pinning models is given by[31-33].

When $X = 7/6$ and $Y = 5/6$,            $\delta T_c$ pinning model

$X = 5/2$ and $Y = -1/2$,            $\delta l$ pinning model

$X = 7/6$ and $Y = -11/6$,           $\delta\varepsilon$ pinning model

As shown in Fig. 7, normalized $J_c(T)$ for SiC + $MgB_2$ composite is well fitted over the main temperature range with the model of $\delta\varepsilon$ pinning (stress/strain field pinning), while nano-C doped $MgB_2$ agrees with the $\delta l$ model and pure $MgB_2$ agrees with the $\delta T_c$ model, both in agreement with our previous reports[34,35]. The nano-C doped $MgB_2$ sample was fabricated by the same processing method as the pure and the SiC composite samples.

In summary, we have succeeded in strain engineering a highly dense $MgB_2$ composite by the thermal diffusion method. In contrast to the well-established C substitution induced enhancement of the superconducting properties, we have demonstrated that the residual thermal stress/strain in SiC-MgB$_2$ composite is caused by the difference in thermal expansion coefficients between MgB$_2$ and SiC, which represents a new mechanism that is responsible for the enhancement in flux pinning and $H_{c2}$ in the SiC-MgB$_2$ composite. XRD and TEM results show that the SiC and MgB$_2$ coexist as two separate but strongly bonded phases. No change in the *a*-axis, a small increase in resistivity and a small decease in $T_c$ as a result of SiC doping indicate a lack of C substitution. The residual thermal strain in the SiC-MgB$_2$ composite is evidenced through the shift of the E$_{2g}$ and PDOS peaks to low frequency, together with the irreversible behavior of the PDOS curves in Raman measurements, the enlargement in the *c*-axis lattice parameter and the increase in micro-strain calculated from Williamson-hall analysis from XRD, the high density of defects on the MgB$_2$ side along the interface as detected by TEM and the behavior of $J_c$ temperature dependence. The present findings have a significant implication, as they open up a new direction for numerous bulk composites that can be strain-engineered to achieve desirable materials properties without significant alteration in intrinsic properties, as compared to chemical substitution.

**ACKNOWLEDGEMENTS**

The authors thank Prof. R. Flükiger, Prof. E.W. Collings, and Dr. T. Silver for their help and useful discussions. This work is supported by the Australian Research Council  (project ID: DP0770205), Hyper Tech Research Inc and the Australian Microcopy & Microanalysis Research Facility (AMMRF) at the Australian Key Centre for Microscopy and Microanalysis at the University of Sydney and the University of New South Wales.

**Figure Captions**

Fig. 1. Rietveld refinement XRD patterns of the pure $MgB_2$ and the 10wt% SiC-$MgB_2$ composite samples that were diffusion reacted at 850°C for 10 hrs, and XRD patterns of the SiC particles added in SiC-$MgB_2$ composite.

Fig. 2. The magnetic $J_c$ versus field at 5K and 20 K for pure and nano-SiC doped samples.

Fig. 3. Upper critical field ($H_{c2}$) and irreversibility field ($H_{irr}$) as functions of the temperature. The inset shows the resistivity of these samples as a function of temperature.

Fig. 4. (a) Plots of the normalized lattice changes for $MgB_2$ and SiC, and the thermal strain in the matrix during cooling from 1123K to 0K. (b) The thermal expansion coefficient ($\alpha$) for $MgB_2$ and SiC as a function of temperature.

Fig. 5. Bright field TEM image for the two-phase composite and defects along the boundaries between the SiC particles and the $MgB_2$ matrix. Defects are indicated by arrows, magnification from (a) 100nm, (b) 10nm , (c) 20nm and (d) 5nm, the (d) HRTEM image of interface and electron diffraction pattern of SiC and $MgB_2$ from each side of the interface, the middle line showing the interface of SiC and $MgB_2$, which showing strong bond of $MgB_2$ with SiC.

Fig. 6. The normalized ambient Raman spectra of samples sintered at 850°C for 10 hrs. The line spectra correspond to measurements taken before, and the dot spectra to measurements taken after cooling to 10K for pure $MgB_2$ (black) and SiC + $MgB_2$ (red) composite.

Fig. 7. The normalized $J_c(T)$ at field $H = 0.1$ T versus reduced temperature ($T/T_c$) and fitting of normalized $J_c(T)$ with various pinning models for pure $MgB_2$, SiC + $MgB_2$ composite, and C substituted $Mg(B_{0.9}C_{0.1})_2$ samples.

**TABLE I. Summary of physical properties of pure and SiC-$MgB_2$ composite samples.**

| Sample | Density | $a$- and $c$-axis parameters | Defect induced non-uniform lattice strain (XRD) | | Mg vacancies | Grain size | $T_c$ | $\rho$ (40K) |
|---|---|---|---|---|---|---|---|---|
| | | | $a$-axis | $c$-axis | XRD | TEM | | |
| | (g/cm$^3$) | (A) | (%) | (%) | (%) | (nm) | (K) | ($\mu\Omega$ cm) |
| **Pure** | 1.86 | 3.085 and 3.5230 | 0.208 | 0.292 | 3.2 | ~100 | 38.4 | 12 |
| **10% SiC** | 1.91 | 3.084 and 3.5282 | 0.306 | 1.13 | 3.6 | ~100 | 37.8 | 16 |

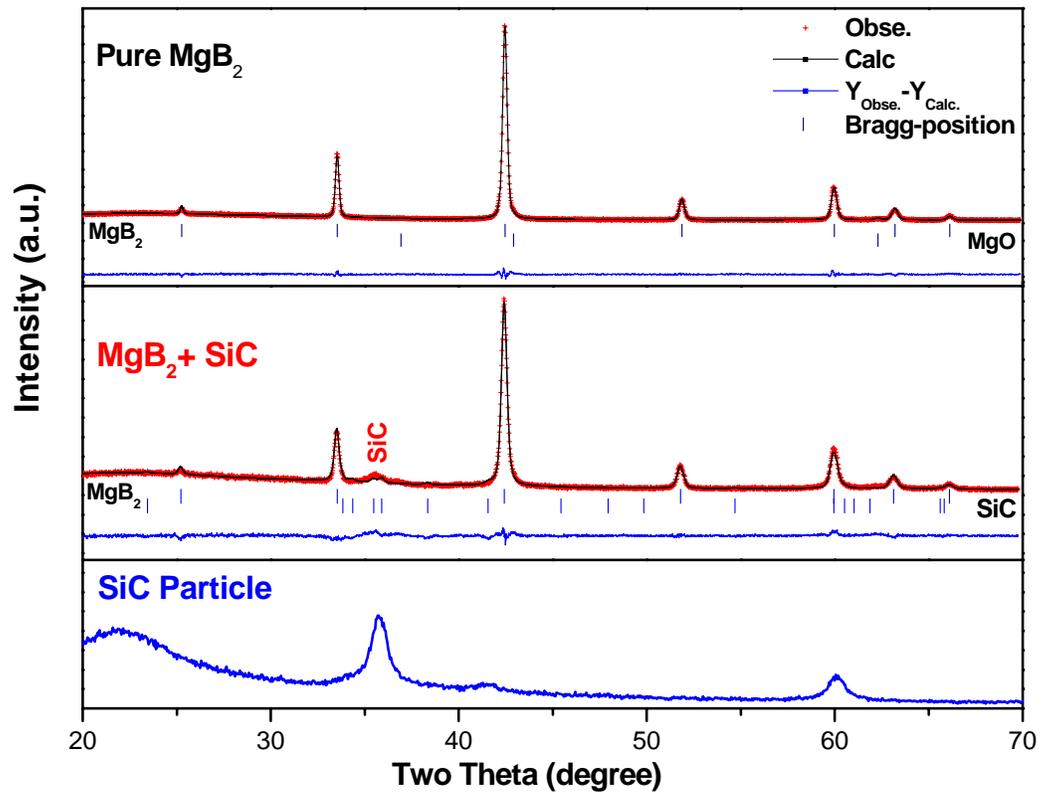

Fig. 1

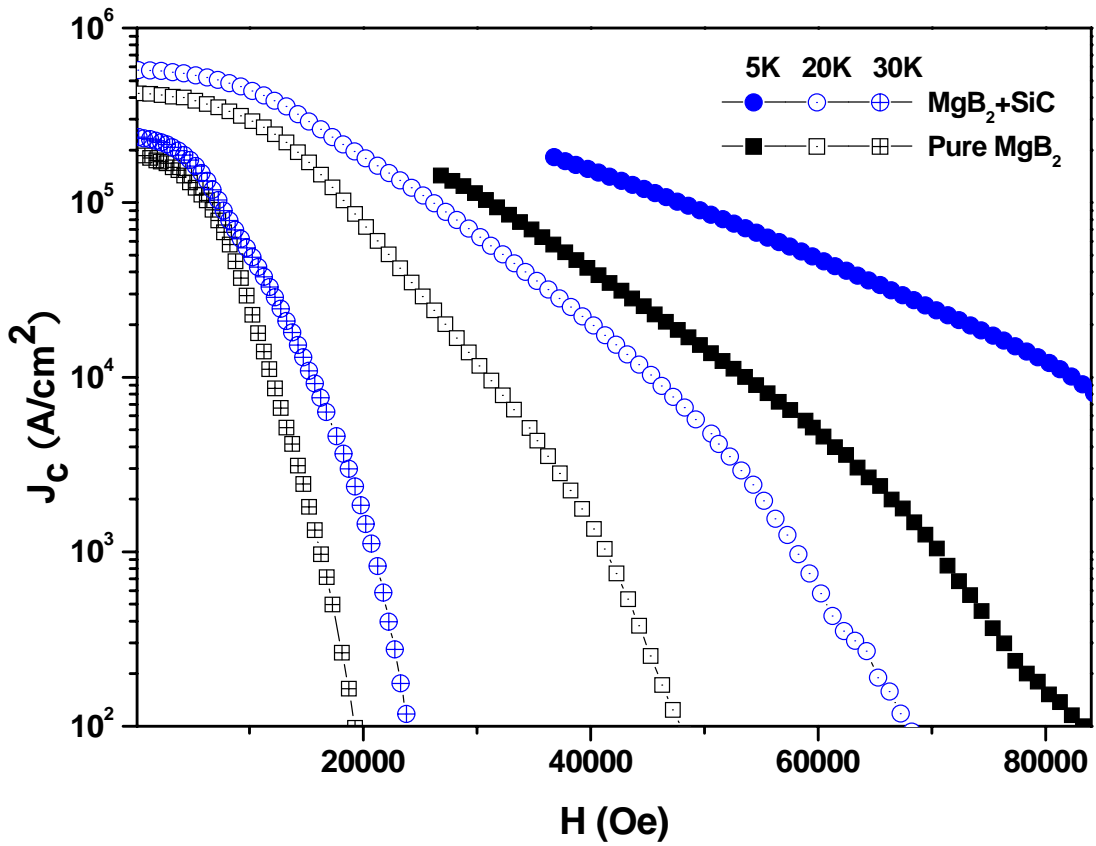

Fig. 2

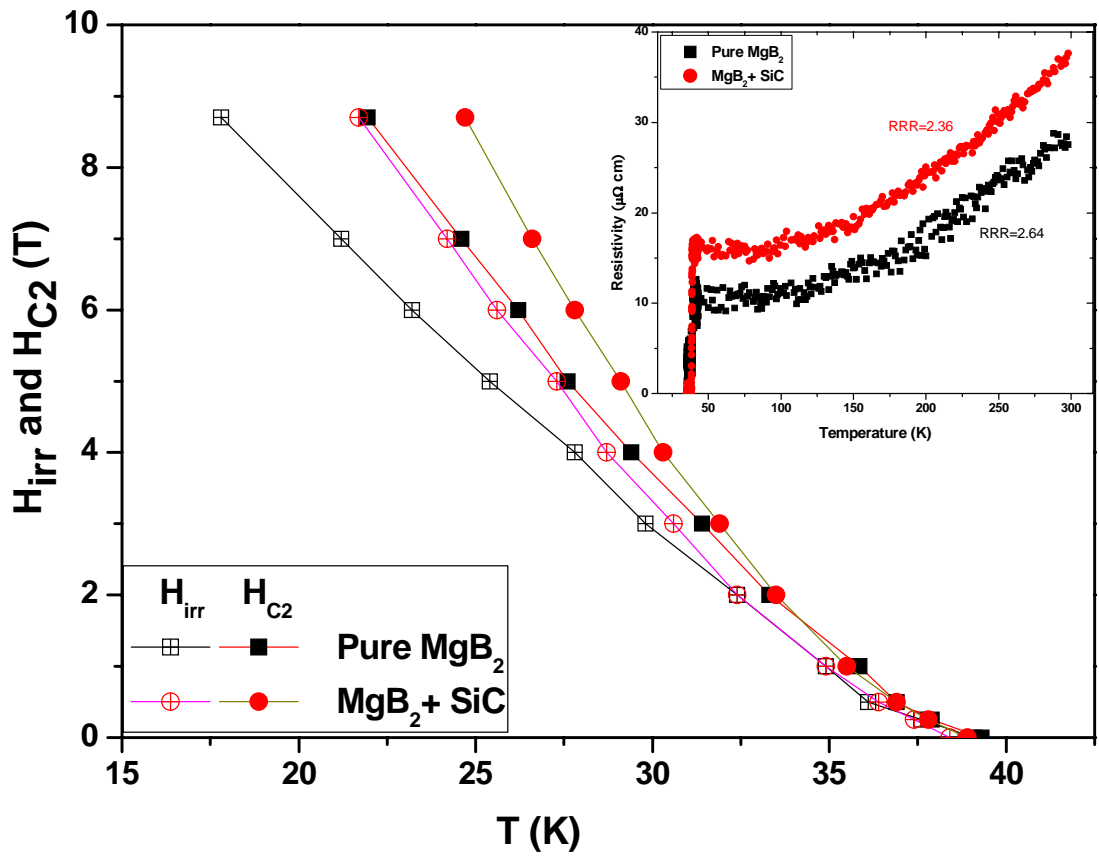

Fig. 3

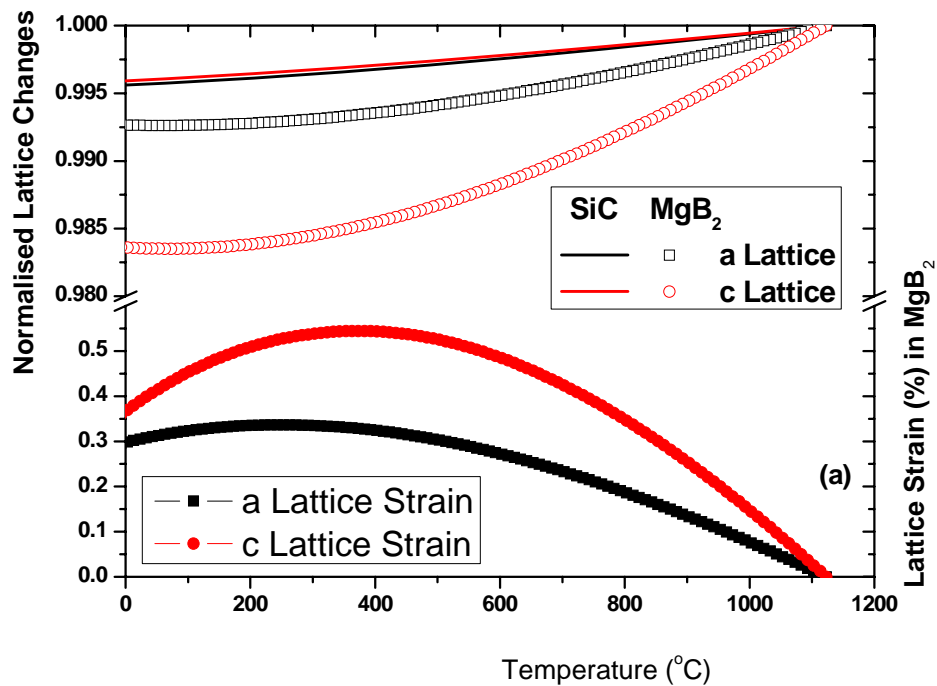

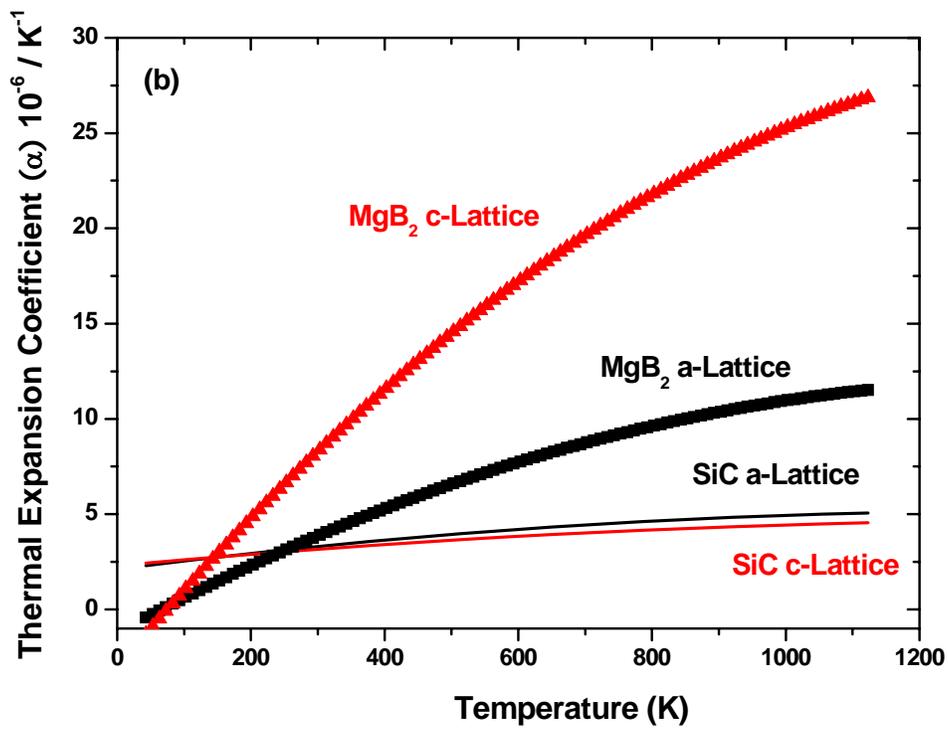

Fig. 4

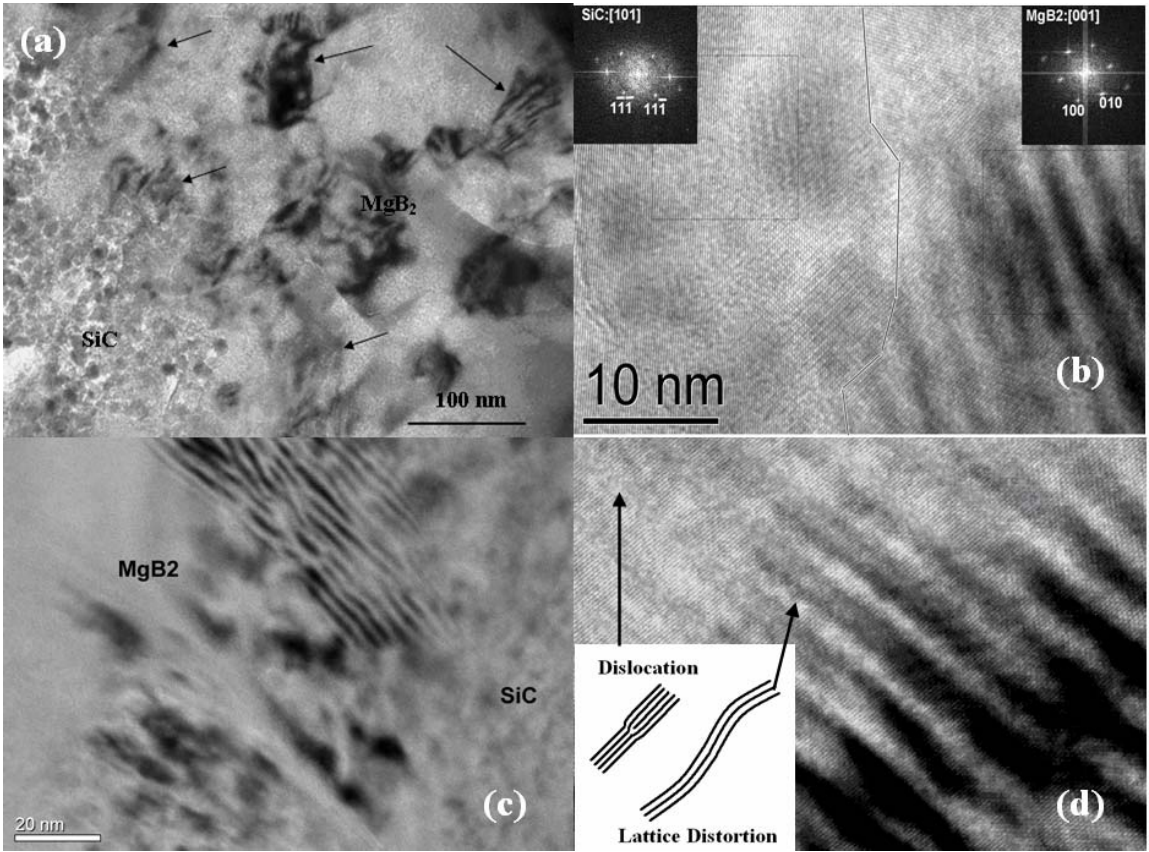

Fig. 5

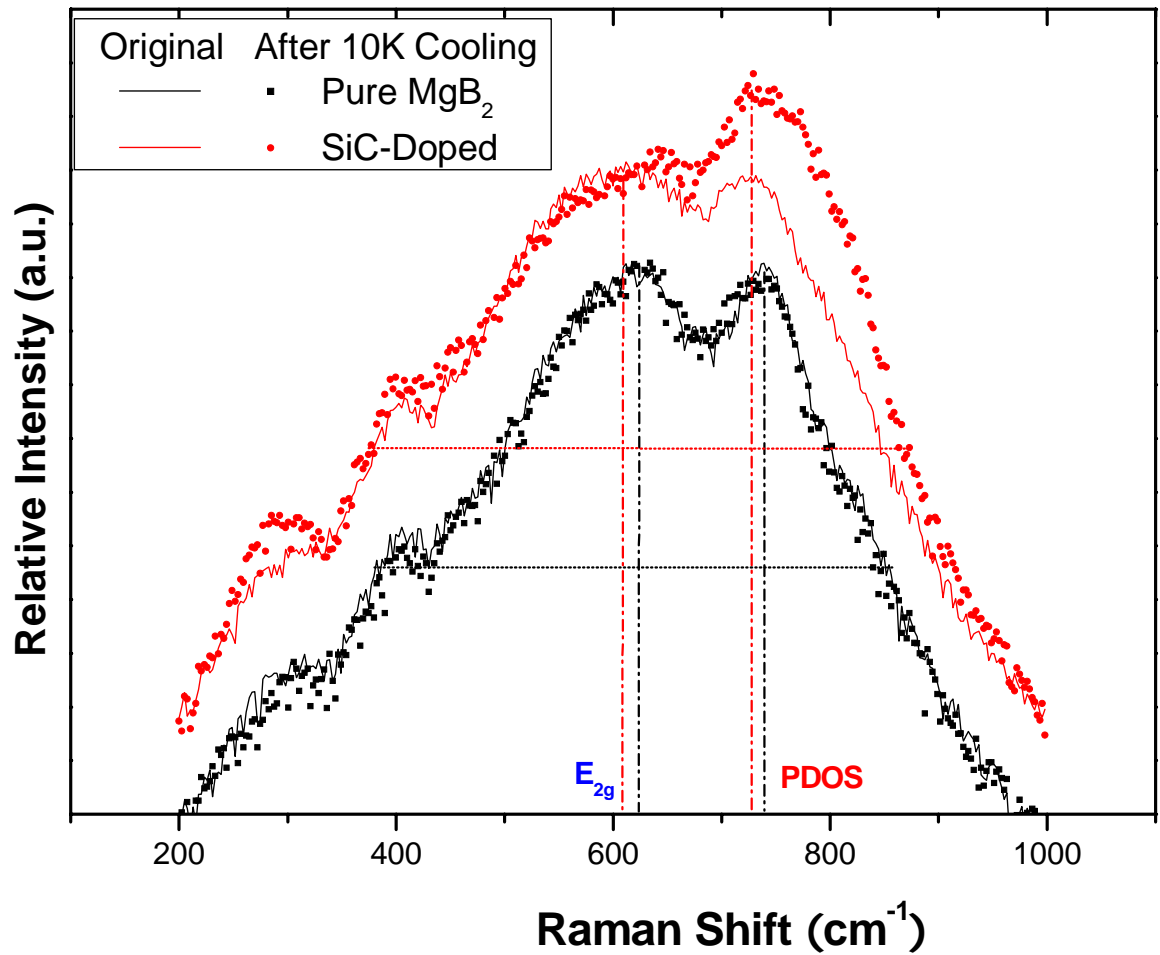

Fig. 6.

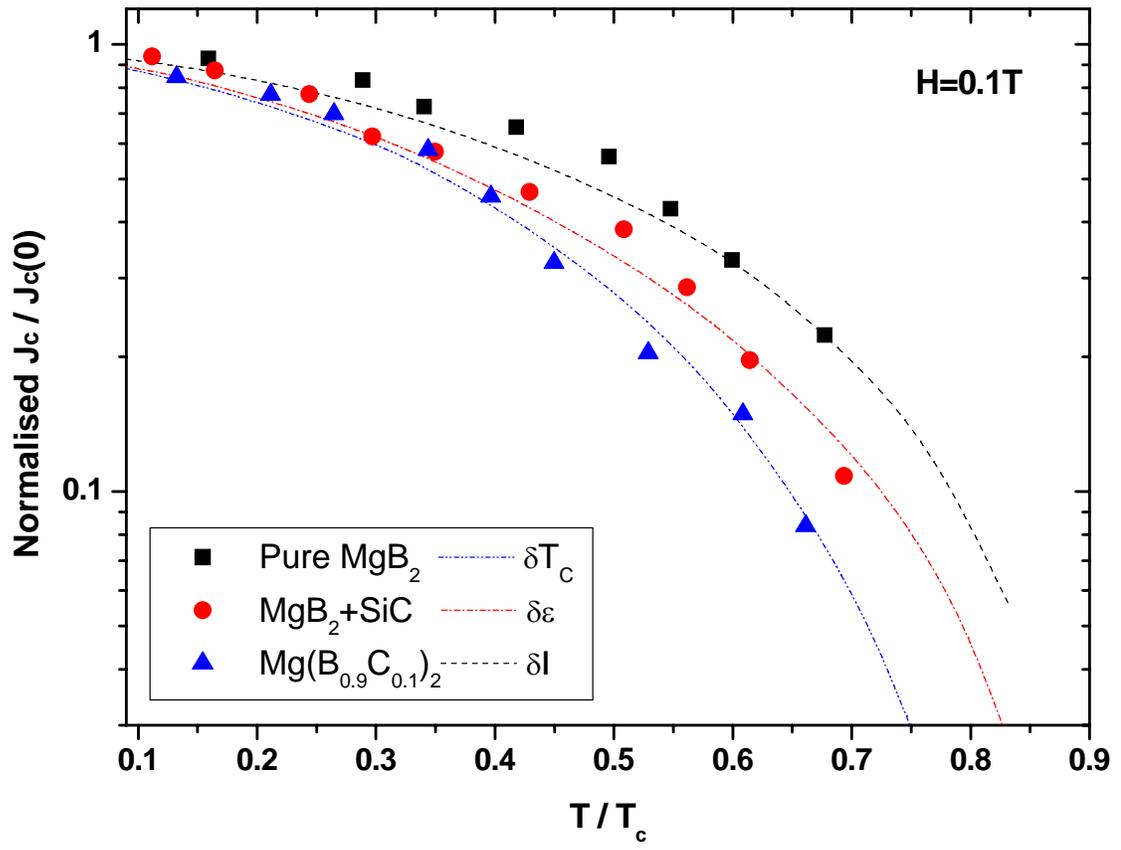

Fig. 7